\documentclass{article}

\usepackage[utf8]{inputenc}
\usepackage{geometry}
\usepackage[numbers,sort&compress]{natbib}
\usepackage{graphicx}
\usepackage{rotating}
\usepackage{amsmath}
\usepackage{amssymb}
\usepackage{bbm}
\usepackage[colorlinks=true, allcolors=blue]{hyperref}
\usepackage{multirow, makecell}
\usepackage{enumerate}
\usepackage{hyperref}
\usepackage{authblk}

\newcommand{\R}[0]{\mathbb{R}}
\newcommand{\ATE}[0]{\text{ATE}}
\newcommand{\ATT}[0]{\text{ATT}}

\newcommand{\prob}[1]{\mathbb{P}\left( #1 \right)}
\newcommand{\esp}[1]{\mathbb{E}\left[ #1 \right]}

\newcommand{\cov}[2]{\text{Cov}\left( #1,\, #2 \right)}
\newcommand{\indi}[1]{\mathbf{1}\left( #1 \right)}
\newcommand{\norm}[1]{\left\lVert#1\right\rVert}
\newcommand{\indep}{\;\raisebox{0.05em}{\rotatebox[origin=c]{90}{$\models$}}\;}

\newcommand{\argmin}[1]{\underset{\substack{#1}}{\text{argmin}}}
\newcommand{\argmax}[1]{\underset{\substack{#1}}{\text{argmax}}}

\newcommand{\keywords}[1]{%
  \begin{center}
    \vspace{0.5em}
    \textbf{Keywords: } #1
  \end{center}%
}

\title{Choosing Covariate Balancing Methods for Causal Inference: Practical Insights from a Simulation Study}

\author[1]{Etienne Peyrot \footnote{Correspondence to: Etienne Peyrot (etienne.peyrot@inserm.fr)}}
\author[1,2]{Rapha\"el Porcher}
\author[1]{François Petit}

\affil[1]{Université Paris Cité and Université Sorbonne Paris Nord, Inserm, INRAE, Center for Research in Epidemiology and StatisticS (CRESS), F-75004 Paris, France}
\affil[2]{Centre d’Épidémiologie Clinique, Assistance Publique-Hôpitaux de Paris,
Hôtel-Dieu, Paris, France}

\date{}

\begin{document}
\maketitle

\begin{abstract}
\textbf{Background}: Inverse probability of treatment weighting (IPTW) is used for  confounding adjustment in observational studies. Newer weighting methods include energy balancing (EB), kernel optimal matching (KOM), and tailored-loss covariate balancing propensity scores (TLF), but practical guidance remains limited. We evaluate their performance when implemented according to published recommendations.\\

\textbf{Methods}: We conducted Monte Carlo simulations across 36 scenarios varying sample size, treatment prevalence, and a complexity factor increasing confounding and reducing overlap. Data generation used predominantly categorical covariates with some correlation. Average treatment effect and average treatment effect on the treated were estimated using IPTW, EB, KOM, and TLF combined with weighted least squares and, when supported, a doubly robust (DR) estimators. Inference followed published recommendations for each method when feasible, using standard alternatives otherwise. An empirical illustration used the \textsc{PROBITsim} dataset.\\

\textbf{Results}: DR reduced sensitivity to the weighting scheme with an outcome regression adjusted for all confounders, despite functional-form misspecification. EB and KOM were most reliable; EB was tuning-free but scale dependent, whereas KOM required kernel and penalty choices. IPTW was more variance sensitive when treatment prevalence was far from 50\%. TLF often traded lower variance for higher bias, producing an RMSE plateau and sub-nominal confidence interval coverage. \textsc{PROBITsim} results mirrored these patterns.\\

\textbf{Conclusions}: Rather than identifying a best method, our findings highlight failure modes and tuning choices to monitor. When the outcome regression adjusts for all confounders, DR estimation can be dependable across weighting schemes. Incorporating weight-estimation uncertainty into confidence intervals remains a key challenge for newer approaches.
\end{abstract}


\section{Introduction}
Since seminal works on propensity scores by Rosenbaum and Rubin \cite{IPW}, causal inference on the effect of treatments using observational data has attracted a lot of attention, both in theoretical \cite{IPTW_norm_weight,Hahn1998} and applied research \cite{Smit2023,Zuo2023}. A key issue with observational data is confounding, and most methods aim to removing, or at least limiting, confounding by balancing confounders between treatment groups.

Among these methods, inverse probability of treatment weighting (IPTW), originally proposed by Robins et al. \cite{AIPW1} building upon survey sampling \cite{IPW1952}, is well-known and commonly used. In parallel, a diverse set of newer weighting strategies has emerged with appealing promises (e.g., model-free or nonparametric behavior), including energy balancing (EB) \cite{Jared2020}, kernel optimal matching (KOM) \cite{kallusATT,kallus2019optimalATE}, and covariate balancing propensity score by tailored loss functions (TLF) \cite{TLF}, among others \cite{Hainmueller2012}. Yet practical guidance for choosing among these techniques is still limited.

Importantly, some of the methods target covariate balance only indirectly (e.g., by minimizing worst-case bias rather than matching moments) \cite{kallusATT,kallus2019optimalATE}. For this reason, it would be ill-advised to appraise these methods by a single balance metric \cite{Franklin2013}. Instead, we assess them by the task that motivates their use in applications: estimating causal effects.

Likewise, in order to avoid favoring any approach and to give each method its best chance, we adopt the estimands and estimators the authors used to present their methods and that recur across the source literature. Accordingly, we focus on the average treatment effect (ATE) and the average treatment effect on the treated (ATT) and, for each method, implement the authors’ common estimators: weighted least squares (WLS) and a doubly robust (DR) estimator. Inference follows the authors’ recommendations where feasible; when those procedures are impractical, for example, when their computational cost is incompatible with the scale of our simulation or when key implementation details are insufficiently specified, we substitute well-documented, standard alternatives and flag the deviation.

Our aim is a pragmatic review rather than a leaderboard. We examine where and why methods exhibit limitations under defensible, good-faith use, for example sensitivity to treatment prevalence and overlap, instability as complexity increases, systematic bias, and difficulties in quantifying uncertainty. We emulate how a practicing biostatistician might proceed, that is, following author guidance, avoiding extensive hyperparameter tuning beyond routine feasibility, and favoring transparent and reproducible choices.

Finally, because several modern methods couple optimization with flexible function classes, constructing valid confidence intervals is nontrivial. We therefore document each method’s recommended inference procedure, for example robust Wald intervals, design-based "honest" bounds, or bootstrap, and when those prescriptions prove impractical or ill-specified in our setting, we explain and justify the alternatives we use.

This paper is structured as follows: we first introduce the weighting methods and the estimators in \autoref{section_settings}. Then, in \autoref{section_simulation_plan}, we detail the data-generative mechanism for the Monte Carlo simulation. In \autoref{section_results}, we present the results of our simulations. In \autoref{section_probit}, we evaluate the methods on \textsc{PROBITsim}, a synthetic observational study built by calibrating a simulated patient cohort to real-world clinical summaries (covariate mix, prevalence, correlations, and overlap) while preserving a known data-generating mechanism. \textsc{PROBITsim} was originally developed as a practice-oriented benchmark for causal-inference methods, providing a testbed with known ground truth; here we use it to verify that the performance patterns seen in the Monte Carlo experiments persist in a clinically realistic setting. Finally, in \autoref{section_discussion}, we conclude with a discussion of the practical limitations and challenges identified in our simulations.

\section{Statistical setting}
\label{section_settings}
\subsection{Causal framework}
We adopt the Neyman-Rubin potential outcomes framework \cite{Rubin1974,Neyman1923}. For each unit, let $Y(0)$ and $Y(1)$ denote the potential outcomes under treatment $A=0$ and $A=1$, respectively, and let $X\in\mathcal{X}\subset\R^p$ be a vector of baseline covariates. Conceptually, each unit is associated with the tuple $(X, Y(0), Y(1), A)$, while only $(X,A,Y)$ is observed. We assume the following:

\begin{itemize}
  \item \textbf{Stable Unit Treatment Value Assumption.} No interference and no hidden versions of treatment; consequently,
  \[
  Y = A\,Y(1) + (1-A)\,Y(0).
  \]
  \item \textbf{Unconfoundedness.} No unmeasured confounding given $X$:
  \[
  \{Y(0),Y(1)\}\ \indep\ A \mid X.
  \]
  \item \textbf{Positivity (weak overlap).} For all $x$ in the support of $X$,
  \[
  0 < \Pr(A=1\mid X=x) < 1.
  \]
\end{itemize}

\subsection{Estimation and notation}
We observe an i.i.d. sample $\{(X_i,Y_i,A_i)\}_{i=1}^n$. We focus on two causal estimands that the original authors used to demonstrate their balancing methods: the average treatment effect (ATE) and the average treatment effect on the treated (ATT).
\[
\ATE := \esp{Y(1)-Y(0)}, \qquad
\ATT := \esp{Y(1)-Y(0)\mid A=1}.
\]
Although ATE is most frequently reported in applied studies, we also evaluate ATT because several of the methods we study were originally developed for ATT and later adapted to ATE \cite{kallusATT,kallus2019optimalATE}. We do not consider the average treatment effect on the controls (ATC), which was not emphasized by the methods under review and is methodologically analogous to the ATT (interchanging treatment and control) and therefore expected to behave similarly.

We denote by $n$ denote the sample size and by $N_0=\sum_i (1-A_i)$ and $N_1=\sum_i A_i$ the size of the control and treated groups respectively.

\subsection{Estimators}
\label{section_estimators}

In this section, we briefly describe the two estimators used in this paper, namely the weighted least squares (WLS) estimator and a doubly robust (DR) estimator. We chose these estimators because they are present in most of the papers presenting the new balancing methods compared in this study. These estimators require a set of weights $(W_i)_i$ computed beforehand by some weighting method. Both estimators can estimate the ATE and ATT assuming the weights also target the same population of interest (i.e. the general population for the ATE or the treated population for the ATT). This section does not present how to get valid confidence intervals for each estimator since the procedure depends on the weighting methods used to obtain the weights. We describe how the balancing methods under evaluation estimate the variance or a confidence interval for these estimators in Section~\ref{section_balancing_methods}.

\subsubsection{Weighted least squares estimator}
\label{subsection_WLS_estimator}
The weighted least squares estimator uses a linear regression of the outcome on the treatment indicator and the intercept. The coefficient assigned to the treatment indicator estimates the treatment effect on the population induced by the weights:
\begin{equation*}
    \esp{Y \mid A} = \alpha + \beta \times A \text{, weighted by } (W_i)_i.    
\end{equation*}
It can be shown that this estimator is exactly the same as the weighted average estimator ($\widehat{\ATE} = \sum_i (2A_i-1) W_i Y_i$) with group-normalized weight (i.e. $\sum_i A_i W_i = \sum_i (1 - A_i) W_i = 1$). If the weights used are those obtained via the IPTW methods, then this estimator is known as the Hájek estimator \cite{hajek1971}.

\subsubsection{Doubly robust estimator}
\label{subsection_DR_estimator}
We considered a doubly robust estimator that relies on the estimation of the response surfaces $\mu_0(X_i) := \esp{Y_i(0) \mid X_i}$ and $\mu_1(X_i) := \esp{Y_i(1) \mid X_i}$. This scheme can yield a tighter estimator than the WLS estimator if the weights correctly balance each group and the response surfaces are well estimated. Moreover, this estimator is consistent as long as one of these two conditions is satisfied.
\begin{align*}
    \widehat{\ATE} &= \sum_{i} \frac{1}{n}\left(\hat{\mu}_1(X_i) - \hat{\mu}_0(X_i)\right) + A_i W_i (Y_i - \hat{\mu}_1(X_i)) - (1 - A_i) W_i (Y_i - \hat{\mu}_0(X_i)),\\
    \widehat{\ATT} &= \frac{1}{N_1} \sum_{i} A_i (Y_i - \hat{\mu}_0(X_i)) - \sum_{i} (1-A_i) W_i (Y_i - \hat{\mu}_0(X_i)).
\end{align*}
with $\hat{\mu}_0$ and $\hat{\mu}_1$ estimations of $\mu_0$ and $\mu_1$ respectively. If the weights used are those obtained via the IPTW methods, then this estimator is known as the augmented inverse probability of treatment weighting (AIPW) estimator \cite{IPTW_norm_weight,Hahn1998,AIPW1,AIPW2,Tsiatis2006}.

\subsection{Balancing methods}
\label{section_balancing_methods}
In this section, we briefly describe the balancing methods compared in our study. We chose as the primary comparison method the inverse probability of treatment weighting \cite{IPW}, to which we compared three recent techniques: energy balancing \cite{Jared2020}, kernel optimal matching \cite{kallusATT,kallus2019optimalATE}, and covariate balancing propensity score by tailored loss function \cite{TLF}.

\subsubsection{IPTW}
\label{subsection_IPW}
IPTW \cite{AIPW1,Tsiatis2006} is a widely used balancing method. Originally developed for estimating the ATE, it assigns weights to each patient based on the inverse probability of receiving the actually assigned treatment. It was later adapted for the ATT. The weights are given by the following formulas:
\begin{align*}
    W^\ATE &= \frac{1}{n}\left(\frac{A}{\hat{e}(X)} + \frac{1 - A}{1 - \hat{e}(X)}\right),\\
    W^\ATT &= \frac{1}{N_1}\frac{\hat{e}(X)}{1 - \hat{e}(X)},
\end{align*}
with $\hat{e}$ an estimation of the propensity score $e$, where $e(X) := \prob{A = 1 \mid X}$.

To derive a confidence interval for the estimators presented in the previous section using these weights, one can rely on M-estimation theory \cite{newey1994, Vaart1998, StefanskiBoos2002} to estimate the standard error with robust (sandwich) standard errors (a.k.a. Huber-White robust standard errors) and build a Wald-type confidence interval. The regularity conditions required by this approach are commonly met in practice. In particular, these conditions are satisfied when the propensity score is estimated by logistic regression.

\subsubsection{Covariate balancing propensity score by tailored loss functions}
\label{subsection_TLF}

Covariate balancing propensity score by tailored loss functions (TLF) \cite{TLF} retains the IPTW approach but fits the propensity model using a loss function tailored to enforce the covariate balance relevant to the target estimand. Concretely, it models the propensity score $e(x)$ with a generalized linear model (GLM) having link $l$, and a prespecified feature set $\mathcal{F}$ for the linear predictor (e.g., main effects, interactions, polynomials, splines). The choice of estimand and link induces a score $S$. The propensity function is then obtained by solving
\begin{equation}
    \label{TLF_general_equation}
    \widehat{e}^{\text{TLF}} = \argmax{p \in \mathcal{P}} \frac{1}{n} \sum_{i = 1}^{n} S\!\left(p(X_i),A_i\right) \;-\; \lambda\, J\!\left(p\right),
\end{equation}
with $\mathcal{P} = l^{-1}(\mathcal{F})$, $J$ a regularizer to limit overfitting, $\lambda$ a parameter that controls the degree of regularization, and $S$ a proper scoring rule whose form depends on the estimand of interest and the GLM link function $l$.

If $l$ is the logit link function, then the expressions of $S$ for the estimands of interest in our study are as follows. Let $(q,a)\in(0,1)\times\{0,1\}$:
\begin{itemize}
    \item ATE: $S(q,a) = (2a-1)\log\!\left(\dfrac{q}{1-q}\right) - \dfrac{a}{q} - \dfrac{1-a}{1-q}$,
    \item ATT: $S(q,a) = (1-a)\log\!\left(\dfrac{1-q}{q}\right) - \dfrac{a}{q}$.
\end{itemize}
The maximizer of  \eqref{TLF_general_equation} estimates the propensity score, from which weights are computed using the standard IPTW formulas, and optionally, group-normalized so that $\sum A_i W_i = \sum (1-A_i)W_i = 1$.

Regarding confidence intervals, the author recommends \cite{TLF} an honest, design-based CI that starts from a usual Wald interval and adds a worst-case bias allowance derived under an RKHS smoothness assumption on the outcome regression. This requires specifying the RKHS and an upper bound on the norm of the outcome regression.

\subsubsection{Energy balancing}
\label{subsection_Energy}
Energy balancing aims at minimizing a discrete version of the energy distance \cite{energyDistance} between the empirical weighted multivariate cumulative distribution function (CDF) of each treatment group and the empirical multivariate CDF of the target population \cite{Jared2020}. This approach generates the following weights:
\begin{align*}
    W^\ATE_{\text{EB}} &= \argmin{\forall i,\, W_i \geq 0\\
    \sum_{i} (1-A_i) W_i = N_0\\
    \sum_{i} A_i W_i= N_1} \mathcal{E}(F_{n,0,w},F_n) + \mathcal{E}(F_{n,1,w},F_n) +
    \mathcal{E}(F_{n,0,w},F_{n,1,w})\label{EB_equation}\tag{2}\\
    W^\ATT_{\text{EB}} &= \argmin{\forall i,\, W_i \geq 0\\
    \sum_{i} (1-A_i) W_i = N_0}  \mathcal{E}(F_{n,0,w},F_{n,1})
\end{align*}
where $F_{n,1}$ (resp. $F_n$) denotes the empirical CDF of the treated group (resp. general population); $F_{n,0,w}$ (resp. $F_{n,1,w}$) represents the weighted empirical CDF of the control group (resp. treated group); and $\mathcal{E}$ is the empirical energy distance on weighted CDF.\\

The third term in the minimization problem for the ATE (\autoref{EB_equation}) is a trick proposed by the authors to enhance the performance of their method by further reducing the heterogeneity of the weighted groups at the price of slightly increasing the distance between each group and the target population. Weights obtained through this balancing method are then group-normalized to satisfy $\sum_i A_i W_i = \sum_i (1-A_i)W_i = 1$.\\
The authors recommend bootstrapping to estimate the confidence intervals.

\subsubsection{Kernel Optimal Matching}
\label{subsection_KOM}
General optimal matching (GOM) aims at finding weights that minimize the worst-case conditional mean-squared error (CMSE) criterion for a treatment-effect estimator over a prespecified class of functions \cite{kallusATT,kallus2019optimalATE}. This min-max view addresses the fact that the true response surfaces are unknown by minimizing a worst-case upper bound on CMSE over plausible class of outcome-regression functions.

Kernel Optimal Matching is an instance of GOM that uses an RKHS as the function class. The advantages of KOM are that an RKHS is flexible enough to be reasonably close to the true response surfaces while resulting in a solvable min-max optimization problem. The general formula to compute weights for the ATE and ATT via KOM is:
\begin{align*}
    W^\ATE_{\text{KOM}} &= \argmin{\forall i,\, W_i \geq 0\\
    \sum_{i} A_i W_i= \sum_{i} (1-A_i) W_i = 1}  W^T \left(\sum_{a\in \{0,1\}}I_a (K_a+\lambda_a I)I_a\right)W - \frac{2}{n}\textbf{1}^T(K_1 I_1 + K_0 I_0)W,\\
    W^\ATT_{\text{KOM}} &= \argmin{\forall i,\, W_i \geq 0\\
    \sum_{i} (1-A_i) W_i = 1} W^T I_0\left(K+\lambda I\right)I_0W - \frac{2}{N_1}\textbf{1}^T I_1 K I_0 W
\end{align*}
where \( K, K_0, K_1 \) are the Gram matrices of the chosen kernels \( \mathcal{K}, \mathcal{K}_0, \mathcal{K}_1 \) respectively, where \( K_{ij} = \mathcal{K}(X_i, X_j) \), and \( I_a = \text{diag}(\indi{A_i = a}) \) with $\indi\cdot$ denotes the indicator function. Finally, $\lambda_0$ and $\lambda_1$ control the trade-off between bias and variance in the control and treated groups respectively.\\

The authors do not provide an unconditional variance estimate for the estimators they developed. Instead, they provide a conditional variance estimate, given the dataset. This implies that the weights, being derived from the covariates, are treated as constants. The procedure to estimate a confidence interval differs between the first paper \cite{kallusATT} and the second paper  \cite{kallus2019optimalATE}. Because the second paper is more recent and generalizes beyond ATT estimation, we adopt its approach. It consists in building a robust Wald CIs from stacked M-estimation \cite{Freedman2006} without accounting for weight uncertainty. The variance of the DR estimator is discussed only in the first article \cite{kallusATT} as the authors do not recommend this estimator in the second article \cite{kallus2019optimalATE} due to practical violations of the positivity assumption and the potential for high bias if the outcome and treatment models are misspecified \cite{Kang2007}.

\section{Simulation study plan}
\label{section_simulation_plan}

Our objective is to provide a pragmatic review of recently proposed balancing approaches alongside IPTW. We aim to characterize the performance these methods achieve when applied as a practicing biostatistician would: in good faith, following the authors’ published recommendations, and without hyperparameter tuning beyond what is typically feasible in routine analyses.

To this end, we use a simple, clinically inspired data-generating mechanism. Covariates are predominantly categorical and exhibit some correlation, reflecting common features of electronic health records and clinical registries. We vary sample size, treatment prevalence (inducing imbalance in sample sizes between treated and control groups), and a general "complexity" factor that strengthens confounding and reduces overlap by increasing the influence of covariates on both treatment assignment and outcome. This design allows us to probe where methods are robust and where they fail, rather than to engineer settings in which any one method excels.

The primary goal is to document practical shortcomings, such as sensitivity to prevalence, instability under higher complexity, and systematic bias, under defensible, author-guided implementations. Beyond characterization, the study serves as decision support by linking failure modes to observable data features (treatment prevalence, overlap, and covariate complexity), it offers practical guidance for method selection under routine analytical constraints. We report performance summaries (RMSE, MAE, bias, variance, and empirical coverage of 95\% confidence intervals) primarily to highlight limits and failure modes, providing a realistic picture of what these methods deliver on data that are simple yet share characteristics of clinical practice.

\subsection{Data-generating mechanism}
The data-generating mechanism described in this section simulates baseline covariates, potential outcomes, and the treatment assignment mechanism. To reflect characteristics of real medical data, the simulated data includes more categorical than numerical covariates and incorporates correlations between covariates. For simplicity, we set a null effect by specifying that $Y(1)$ and $Y(0)$ share the same conditional distribution given $X$; consequently, the true ATE and ATT are $0$ and makes bias comparisons across scenarios straightforward. \\

For each patient, the first step is to generate a Gaussian vector $\widetilde{X}=(\widetilde{X}^{(1)}, \widetilde{X}^{(2)}, \dots, \widetilde{X}^{(10)})$ drawn from a multivariate Gaussian distribution with mean zero and a specified covariance matrix. Each component of the Gaussian vector has variance $1$, and the covariance is set to zero for all pairs of variables except for the following:
\begin{align*}
\cov{\widetilde{X}^{(1)}}{\widetilde{X}^{(5)}} = \cov{\widetilde{X}^{(3)}}{\widetilde{X}^{(8)}} = 0.2\,,\\ 
\cov{\widetilde{X}^{(2)}}{\widetilde{X}^{(6)}} = \cov{\widetilde{X}^{(4)}}{\widetilde{X}^{(9)}} = 0.9\,.
\end{align*}

The following transformation is applied to $\widetilde{X}$ to get binary covariates:
\begin{equation*}
    X^{(i)} =
    \begin{cases}
        \widetilde{X}^{(i)} \quad \textnormal{if} \quad i \in \{2,4,7,10\},\\
        \indi{\widetilde{X}^{(i)} > 0} \quad \textnormal{otherwise}.
    \end{cases}
\end{equation*}

Given the data-generating mechanism for $X$, we generate the potential outcomes $Y(0), Y(1)$ and the treatment assignment mechanism $A$ as follows:
\begin{align*}
    Y(0) \text{ and } Y(1) &\sim \text{Bernoulli}\left\{ \text{logit}^{-1}\left[a_0 + g \left(X a + \frac{1}{2} X^{(3)} {X^{(4)}}^2\right) \right] \right\},\\
    A &\sim \text{Bernoulli}\left\{ \text{logit}^{-1}\left[ b_0 + g \left(X b + \frac{1}{2} X^{(1)} {X^{(2)}}^2\right) \right] \right\}
\end{align*}
where
\begin{align*}
    a &= ( 0.9,\, -1.08,\, - 2.19,\, -0.6,\, \phantom{-}0\phantom{.0},\, \phantom{-}0\phantom{.0},\, 0\phantom{.0},\, 0.71,\, -0.19,\, 0.26 )^\top,\\
    b &= ( 0.8,\, -0.25,\, \phantom{-}0.6\phantom{0},\, -0.4,\, -0.8,\, -0.5 ,\, 0.7,\, 0\phantom{.00},\, \phantom{-}0\phantom{.00},\, 0\phantom{.00})^\top.
\end{align*}
The constants $a_0$ and $b_0$ control the number of events and the proportion of treated respectively. The variable $g$ affects both the probability of receiving the treatment and the potential outcomes by scaling the parts controlled by baseline covariates in each formula. Thus, $g$ has a direct effect on confounding bias and overlap.\\

From this data-generating mechanism, we created several scenarios by calibrating $a_0$, $b_0$ and $g$ in order to:
\begin{enumerate}
    \item set the probability of an event occurring (i.e. $\prob{Y=1}$) to $25\%$ in all scenarios;
    \item create scenarios with low, moderate, and high proportions of treated with $\prob{A=1}$ set to $25\%$, $50\%$, and $75\%$ respectively;
    \item create scenarios with a low, moderate, and high level of complexity with the bias of the crude estimator of the ATE (i.e. $\esp{N_1^{-1}\sum_i A_i Y_i - N_0^{-1}\sum_i (1-A_i) Y_i}$) equals to $0.05$, $0.10$, and $0.15$ respectively, and an overlap (ie. $\prob{5\% \leq e(X)\leq95\%}$) equals to $99.5\%$, $95\%$, and $75\%$ respectively for a proportion of treated of $50\%$.
\end{enumerate}

The exact values of $a_0$, $b_0$, and $g$ for each scenario are provided in the Supplementary Materials. For more details on how these parameters affect the distribution of data, an R Shiny app is available in the GitHub repository for this article \citep{benchmark_repo}.\\

To assess how balancing methods perform with smaller sample sizes, we created scenarios with 250, 500, 1000, or 2000 observations. In total, there are 36 scenarios defined by sample size (250, 500, 1000, or 2000), proportion of treated ($25\%$, $50\%$, or $75\%$), and the level of complexity (low, moderate, or high).

\subsection{Implementation of balancing methods and estimators}
\label{subsection_method_implementation}
This section provides implementation details for the balancing methods  described in Section~\ref{section_balancing_methods} and the estimators introduced in Section~\ref{section_estimators}, including the practical choices required to apply them and the software used. All the code for this study is available at this article’s GitHub repository \cite{benchmark_repo}.

\paragraph{Estimators}
We implemented the WLS estimator in base R. For the DR estimator, we first fit outcome models within each treatment arm. Because the outcome is binary in our data-generating mechanism, we used logistic regression including all covariates, with no variable selection or feature engineering. Confidence intervals were derived via M-estimation from a stacked system of estimating equations: under EB and KOM weights, we stacked the estimating equation for the target estimand with the outcome-regression score functions; for IPTW and TLF, we additionally included the propensity-score model score functions. We implemented these procedures in base R.

\paragraph{IPTW}
We estimated propensity scores via logistic regression including all covariates (no variable selection or feature engineering) and implemented the weighting in base R. Wald-type confidence intervals were constructed using robust sandwich standard errors, implemented in base R.

\paragraph{KOM}
KOM requires choosing, for each treatment arm $a \in \{0,1\}$, a kernel $\mathcal{K}_a$ and a bias-variance penalty $\lambda_a$. Once the kernels were chosen, we tuned the kernel hyperparameters following the guidance in \cite[\S3.6]{kallus2019optimalATE}, within each arm, by maximizing the Gaussian process (GP) marginal likelihood for the outcome and set $\lambda_a$ to the noise-to-signal ratio. We used a Gaussian kernel for both arms. While \cite[\S3.6]{kallus2019optimalATE} suggests a polynomial Mahalanobis kernel as a general default and lists Gaussian/Matérn as alternatives, we adopted a $C_0$-universal kernel in light of broader recommendations for KOM in \cite[\S4.7]{kallusATT}. Moreover, the polynomial Mahalanobis kernel requires selecting an integer degree $d$ (among other hyperparameters), which makes automated tuning less convenient than for the Gaussian kernel, whose primary length scale is a continuous parameter optimized by marginal likelihood.

To the best of our knowledge, there is no dedicated R package on CRAN implementing KOM. We therefore relied on the author's reference R code from a GitHub repository developed for a related article \cite{kalluscodearticle}, with the following modifications: (i) we modified the computation of $\lambda$ to match the recommendation of \cite[\S3.6]{kallus2019optimalATE}; (ii) we replaced the quadratic-programming (QP) solver \texttt{Gurobi} \cite{gurobi} with \texttt{OSQP} \cite{osqp}, an open-source solver suited to large convex QPs; and (iii) we added analytic derivatives of the GP marginal likelihood to accelerate hyperparameter selection. We followed this strategy (see Section~\ref{subsection_KOM}) and implemented the procedure in base~R.

\paragraph{TLF}
We implemented the TLF method following \cite{TLF} and the author’s R source package \texttt{covalign} \citep{covalignR}. In line with the practical guidance for low-dimensional covariates, we modeled the propensity score using a Laplacian kernel and selected the regularization parameter $\lambda$ by cross-validation (\S5.6), minimizing the average norm of the tailored-loss gradient on validation folds. The author recommends in \cite[\S7]{TLF} to use universal kernels, particularly the Laplacian, and encourages trying multiple kernels or bandwidths as a sensitivity analysis; no bandwidth-selection rule is prescribed. Accordingly, we fixed the kernel family to Laplacian and set its bandwidth using the "median heuristic" (or its inverse, depending on parameterization), a standard convention in kernel methods \citep{gretton12,garreau2018,Ramdas2015,Schrab2023}. To reduce computation, we pre-tuned $\lambda$ per scenario and estimand via a grid over 50 replicated datasets, choosing the value that minimized the average imbalance proxy advocated in \cite[\S5.6]{TLF}.

The author recommends an honest, design-based CI that starts from a Wald interval and then adds a "worst-case bias" margin based on a smoothness assumption for the outcome in a specified RKHS. In practice, this requires (i) choosing that function space and an upper bound on the outcome’s complexity, which cannot be verified from the data and for which the paper provides limited implementation guidance in low or moderate dimensions; (ii) accepting additional restrictions (for non-ATT estimands, a constant treatment effect); and (iii) tolerating intervals that can be overly conservative and therefore uninformative. For these reasons, we did not adopt this CI and instead used the same approach as for IPTW: robust (sandwich) standard errors obtained from a stacked M-estimation system that includes the propensity-score model.

\paragraph{EB}
Energy balancing requires no user-specified tuning. We computed EB weights using the R package \texttt{WeightIt} with \texttt{method="energy"} \cite{WeightItR}. Although the original proposal recommends bootstrap-based confidence intervals \cite{Jared2020}, full resampling was computationally prohibitive for our simulation grid. Instead, we constructed Wald-type intervals using the same M-estimation sandwich procedure as for KOM (stacking the estimating equation for the estimand and additionally, for DR, the outcome-regression score functions). This was implemented in base~R.

\subsection{Performance metrics}
\label{subsection_performance_metrics}
We compared estimators using five Monte Carlo performance metrics: root mean squared error (RMSE), mean absolute error (MAE), empirical variance, empirical bias, and confidence interval coverage. Let $\widehat\theta^{(i)}$ denote the estimate from replication $i$ $(i=1,\dots,R)$ and $\overline{\widehat\theta}=R^{-1}\sum_{i=1}^R \widehat\theta^{(i)}$. Then
\[
\mathrm{RMSE}(\widehat\theta)=\sqrt{\frac{1}{R}\sum_{i=1}^R\bigl(\widehat\theta^{(i)}-\theta\bigr)^2},\qquad
\mathrm{MAE}(\widehat\theta)=\frac{1}{R}\sum_{i=1}^R\bigl|\widehat\theta^{(i)}-\theta\bigr|,
\]
\[
\mathrm{Bias}(\widehat\theta)=\overline{\widehat\theta}-\theta,\qquad
\mathrm{Var}(\widehat\theta)=\frac{1}{R-1}\sum_{i=1}^R\bigl(\widehat\theta^{(i)}-\overline{\widehat\theta}\,\bigr)^2,
\]
\[
    \text{Coverage}(\hat{\theta}) = \frac{1}{R} \sum_{i=1}^R \indi{\theta \in \text{CI}_{95\%}(\hat{\theta}^{(i)})}.
\]
\section{Results}
\label{section_results}

In this section, we present the simulation results. All simulations were conducted in R (version 4.2.1; \cite{R}). For each of the 36 scenarios, we generated 5{,}000 datasets. For each dataset, we applied the four weighting methods and, for each estimator, computed a treatment-effect estimate using the resulting weights. The 5{,}000 estimates per scenario--method--estimand combination were then used to compute the performance metrics described in Section~\ref{subsection_performance_metrics}. Detailed results are provided in the Supplementary Materials. To facilitate exploration, we provide an R Shiny application at the article’s GitHub repository \citep{benchmark_repo}.

We first report results by estimand, followed by empirical coverage of the 95\% confidence intervals.

\subsection{ATE}
\begin{figure}[htbp]
    \centering
    \includegraphics[width=\linewidth]{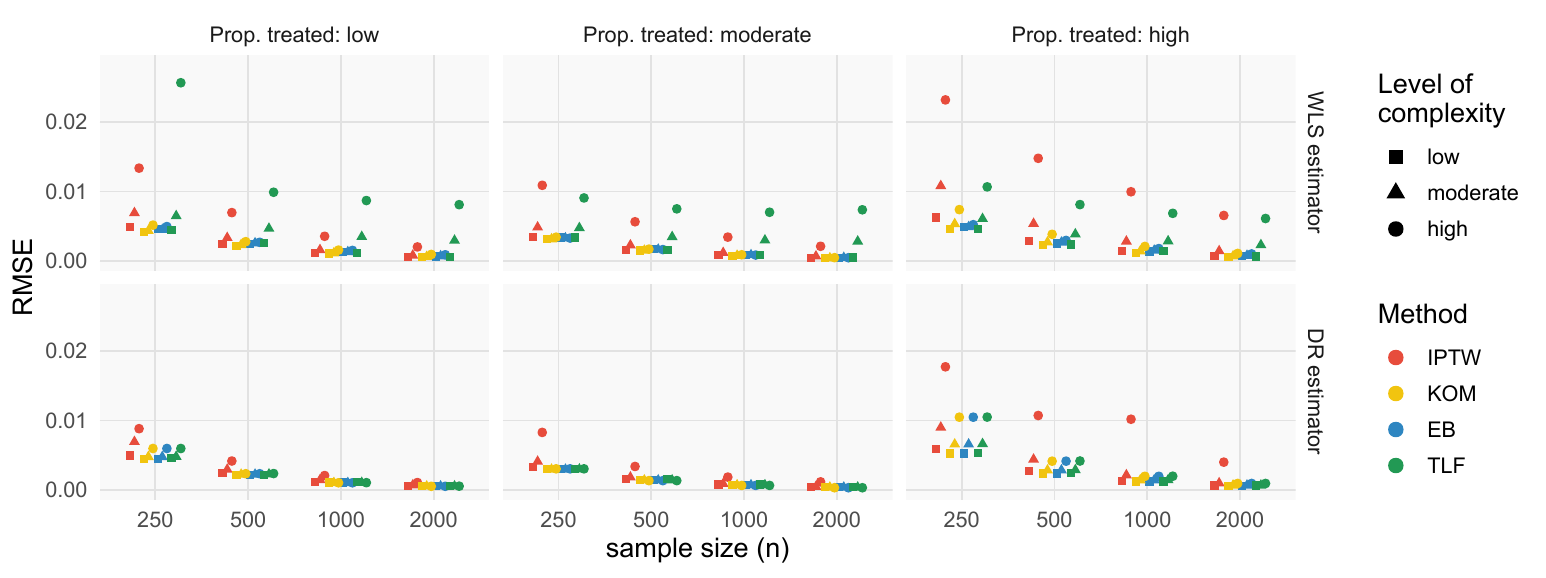}
    \caption{RMSE of the ATE estimate for each weighting method-estimator couple and scenario.}
    \label{fig:simulation_res_ATE}
\end{figure}
\autoref{fig:simulation_res_ATE} summarizes ATE performance in terms of RMSE. Across designs, RMSE decreases as sample size increases and grows with scenario complexity (i.e., increased confounding and reduced overlap); it is smallest under moderate treatment prevalence and larger when the proportion treated is low or high.

Overall, the DR estimator yields lower RMSE than WLS. There are isolated cases within specific methods where WLS is slightly better, but these differences are minor.

Turning to weighting methods, EB and KOM deliver similar and comparatively low RMSE across scenarios and show relatively high robustness to the complexity of the data-generating mechanism. In contrast, TLF exhibits a clear RMSE plateau as $n$ increases, so its RMSE is generally higher than for the other methods. In the most challenging settings, IPTW performs worst, and like TLF, its RMSE increases substantially as scenario complexity increases.

Supplementary diagnostics (bias, variance, and MAE) clarify these patterns. EB has slightly lower bias than KOM but slightly higher variance, resulting in nearly identical RMSE. TLF shows very low variance together with comparatively high bias, consistent with its RMSE plateau. IPTW displays the largest variance and greater sensitivity to scenario complexity, which largely explains its poorer RMSE. Overall, the RMSE ordering is driven mainly by variance, except for TLF, where bias plays a larger role.

\subsection{ATT}
\begin{figure}[htbp]
    \centering
    \includegraphics[width=\linewidth]{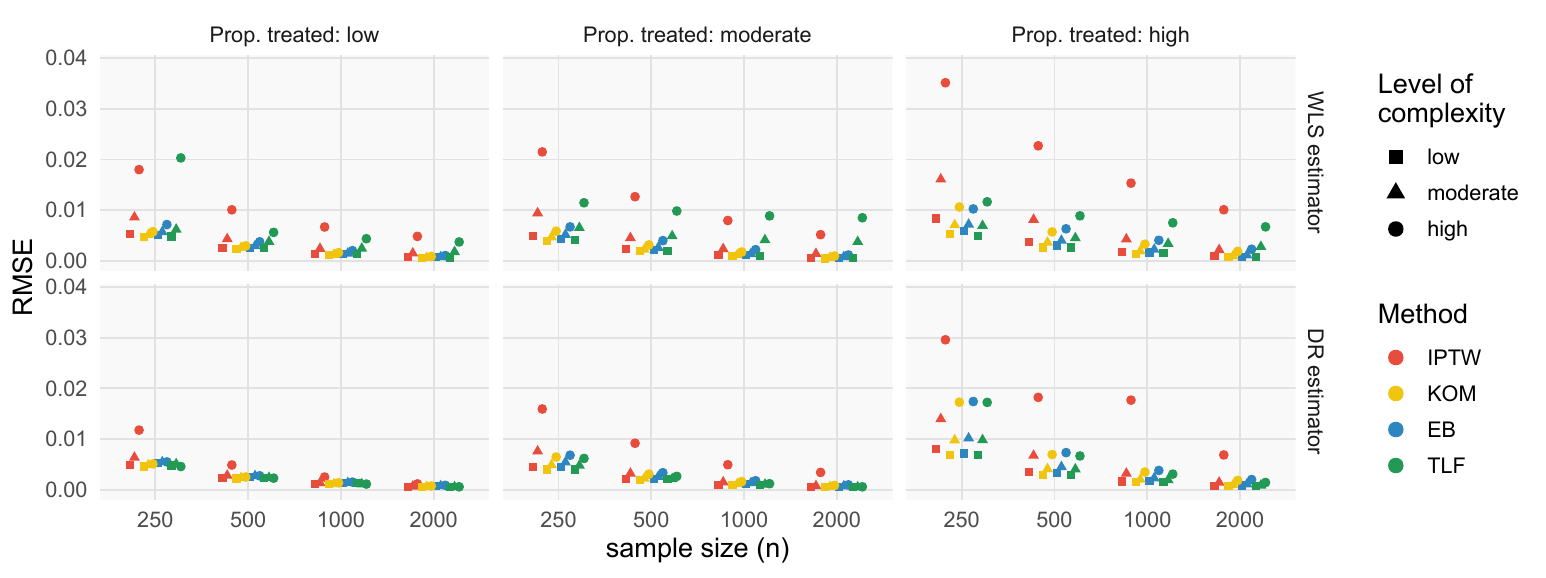}
    \caption{RMSE of the ATT estimate for each weighting method-estimator couple and scenario.}
    \label{fig:simulation_res_ATT}
\end{figure}

Most observations made for the ATE remain true for the ATT; here we highlight the differences shown in \autoref{fig:simulation_res_ATT} and the Supplementary Figures. First, the dependence on treatment prevalence is more asymmetric: RMSE is lowest when the proportion treated is high and degrades notably when it is low, reflecting that ATT targets the treated population. Overall, the DR estimator again outperforms WLS, with only minor, isolated cases within specific methods where WLS is slightly better.

By method, EB and KOM continue to yield similarly low RMSE, though this time KOM seems to have a slight advantage over EB, and remain comparatively robust to increased scenario complexity. TLF shows the same RMSE plateau as sample size grows that is already present for the ATE, so its RMSE is generally higher than for the other methods. In the more challenging designs, IPTW performs worst and its RMSE increases as complexity rises, especially when the proportion treated is low.

Supplementary diagnostics (bias, variance, and MAE) indicate that EB still has slightly lower bias than KOM but a higher variance.  TLF combines very low variance with comparatively high bias, consistent with its RMSE plateau. IPTW exhibits the largest variance and the greatest sensitivity to complexity, which largely explains its poor RMSE. As with ATE, the RMSE ordering is primarily driven by variance, with TLF being the main exception where bias plays a larger role.

\subsection{Coverage}

\begin{figure}[htbp]
    \centering
    \includegraphics[width=\linewidth]{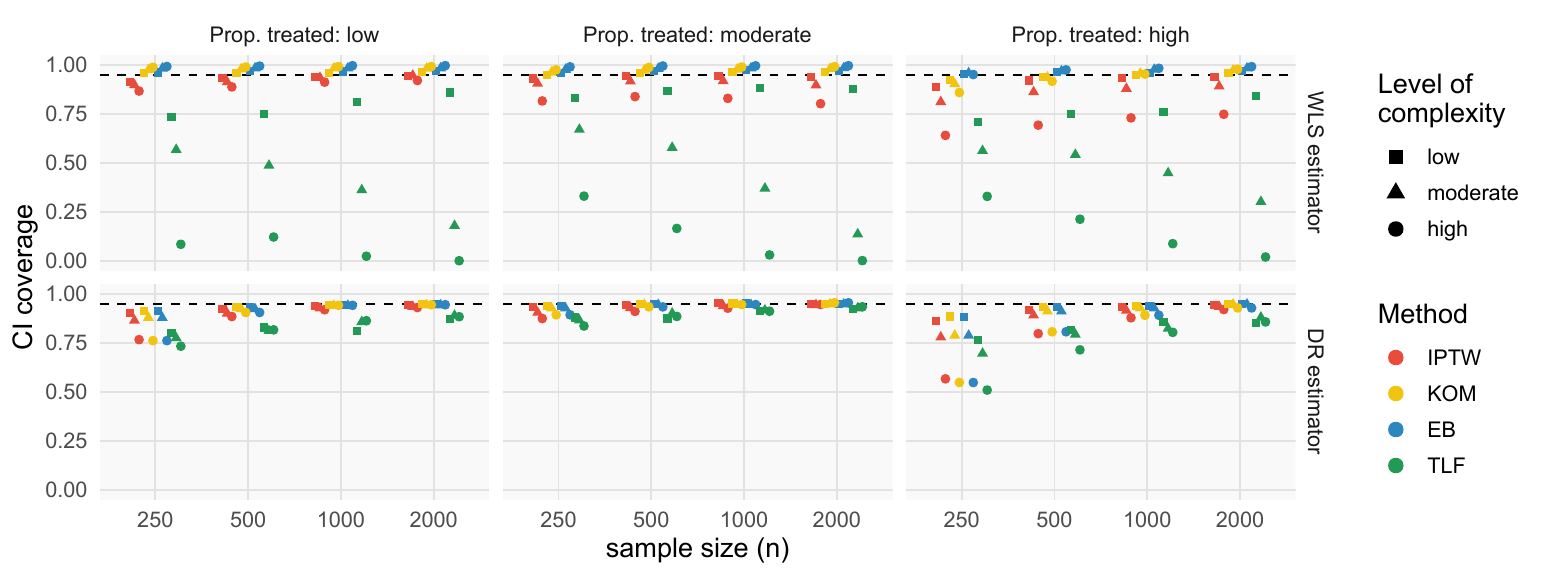}
    \caption{Coverage of the $95\%$ confidence interval of ATE estimate for each weighting method-estimator couple and scenario. The black dashed line is the nominal coverage ($95\%$).}
    \label{fig:simulation_coverage_ATE}
\end{figure}
\autoref{fig:simulation_coverage_ATE} shows empirical coverage of 95\% confidence intervals. Coverage generally improves with sample size and degrades as scenario complexity increases. EB and KOM achieve coverage closest to nominal in most settings, especially with the DR estimator at moderate and large $n$. For these methods, WLS tends to over-cover, whereas DR tends to under-cover. IPTW typically under-covers but can be near nominal in low-complexity scenarios; its coverage deteriorates as complexity increases compared with EB and KOM. TLF performs worst overall, consistent with its higher bias; as $n$ increases, coverage with WLS declines and DR shows only limited gains relative to the other methods.

For the ATT (see Supplementary Figures), patterns broadly align with those for the ATE, with two notable differences. First, under DR, coverage increases as treated prevalence decreases; at moderate and large $n$, EB and KOM attain near-nominal coverage when the treated proportion is small. Second, under WLS, coverage decreases for IPTW, EB, and KOM, but increases for TLF when treated prevalence is low. This results in a correction of the mild over-coverage of EB and KOM, bringing them closer to nominal coverage, and a worsening of IPTW’s coverage, particularly at low treated prevalence. Despite the gains under WLS, TLF remains the weakest performer overall.

\section{Application to PROBITsim data}
\label{section_probit}

In this section, we assess the methods on \textsc{PROBITsim} \cite{Goetghebeur2020}. \textsc{PROBITsim} is a synthetic observational cohort, calibrated to summary characteristics of the \textsc{PROBIT} randomized trial, but not a direct replication. It generates individual mother-infant pairs and focuses on infant weight at 3 months as the primary outcome. The data include baseline covariates commonly available in clinical records (maternal age; region: urban/rural, west/east; education: low/intermediate/high; allergy history; smoking during pregnancy) and birth-related variables (infant sex, birthweight, caesarean section). The sample size is set to $n=17{,}044$, as in PROBIT and the data-generating mechanism induces realistic confounding and selected interactions (e.g., between breastfeeding duration and education, smoking, or birthweight). 
\textsc{PROBITsim} was created as a practice-oriented benchmark for causal-inference methods: by calibrating to real clinical summaries (covariate mix, prevalences, correlations, and overlap) while preserving a known data-generating mechanism, it delivers trial-like realism with a known ground truth for the causal estimands. Treatment assignment is intentionally nonrandom to induce realistic confounding and overlap patterns, making it a useful testbed to verify whether the patterns observed in our Monte Carlo study persist under clinically plausible conditions.\\

The intervention is represented as a chain of four linked point exposures with the temporal order:
\begin{itemize}
    \item[(i)] $A_1$, offer of a breastfeeding encouragement program (BEP);
    \item[(ii)] $A_2$, BEP uptake;
    \item[(iii)] $A_3$, breastfeeding initiation;
    \item[(iv)] $A_4$, breastfeeding maintained for 3 months.
\end{itemize}
 Importantly, BEP uptake $A_2$ is defined as conditional on being offered the BEP $A_1$: a mother can enroll only if an offer was made.
For each unit, \textsc{PROBITsim} generates potential outcomes under alternative exposure strategies, making the true treatment effects known and allowing us to evaluate estimators against ground truth. For BEP uptake ($A_2$), the focus of our analysis, the true treatment effects are $\ATE=165$~g and $\ATT=153$~g for weight at 3 months. Code and further details are available from the authors’ materials.\\

We applied the same estimators as in the simulation study; the only change was to use linear regression for the outcome models, since the outcome is continuous. For IPTW, the propensity score model followed the authors’ specification for exposure $A_2$. Implementations of TLF, EB, and KOM followed the strategies detailed in Section~\ref{subsection_method_implementation} without further modification. Code for the analysis is available on this article GitHub repository \citep{benchmark_repo}.

\begin{table}[tbhp]
    \centering
    \caption{ATE and ATT estimates (with 95\% CIs) for the effect of BEP participation ($A_2$) on infant weight (g) at 3 months in \textsc{PROBITsim}. True effects: ATE = $165$~g, ATT = $153$~g.}
    \label{table:probitsim_res_1}
    \begin{tabular}{lcc|cc}
        \hline
         & \multicolumn{2}{c|}{ATE} & \multicolumn{2}{c}{ATT} \\
        Method & WLS & DR & WLS & DR \\
        \hline
        IPTW & $165\,[146, 184]$ & $164\,[145, 183]$ & $148\,[129, 167]$ & $149\,[130, 167]$ \\
        TLF  & $188\,[170, 207]$ & $165\,[146, 184]$ & $171\,[153, 190]$ & $148\,[130, 167]$ \\
        EB   & $158\,[138, 178]$ & $165\,[146, 184]$ & $148\,[128, 167]$ & $148\,[128, 167]$ \\
        KOM  & $167\,[148, 186]$ & $165\,[146, 184]$ & $149\,[131, 168]$ & $149\,[130, 167]$ \\
        \hline
    \end{tabular}
\end{table}

The estimates and confidence intervals in \autoref{table:probitsim_res_1} show that, across methods, the DR ATE estimates align closely and center on the true value ($165$~g) with similar interval widths, indicating that the outcome model largely drives performance in this application. In contrast, WLS reveals method-specific differences. Under WLS, TLF is an outlier: it overestimates the ATE ($188$~g; 95\% CI $[170, 207]$) and does not include the true value, whereas EB, KOM, and IPTW all include $165$~g (EB slightly low at $158$~g; KOM slightly high at $167$~g; IPTW at $165$~g). For the ATT, all methods’ CIs contain the true value ($153$~g); EB/KOM/IPTW with WLS cluster around $148-149$~g (slightly low), whereas TLF-WLS is higher ($171$~g) with the lower CI bound at $153$~g. Precision is broadly similar across methods, so differences are primarily in centering rather than interval width. Overall, these empirical results are consistent with the simulation study: DR tends to be more reliable across weighting choices, EB and KOM behave stably under WLS, IPTW performs reasonably in this setting, and TLF exhibit noticeable bias when not paired with the DR estimator.

\section{Discussion}
\label{section_discussion}
We emphasize practical implications of the results. First, the DR estimator consistently reduces sensitivity to the choice of weights, though this remains contingent on a reasonably specified outcome regression. Second, EB and KOM behave similarly under authors' guidance: EB is tuning-free but scale-dependent (standardization matters), whereas KOM requires kernel and $\lambda$ choices; following the authors’ recommendations yielded stable performance here. Third, TLF exhibits very low variance but comparatively higher bias, leading to an RMSE plateau and sub-nominal coverage. Fourth, IPTW is workable but variance-sensitive, particularly as complexity rises.\\

\paragraph{On the relation between EB and KOM.}
Although EB is presented as a distance-based method, it can be seen as an instance of KOM. Indeed, EB chooses weights by minimizing a weighted empirical energy distance between covariate distributions, where the underlying dissimilarity is the squared Euclidean norm $\norm{\cdot}_2^2$ on $\mathbb{R}^p$ \cite{Jared2020}. This energy distance is equal (up to a constant factor) to the squared maximum mean discrepancy (MMD) computed in a RKHS induced by the following distance-induced kernel \cite{Sejdinovic2013} $k(x,x') = \tfrac12\bigl(\norm{x-x_0}_2^2 +\norm{x'-x_0}_2^2 - \norm{x-x'}_2^2\bigr)$, for any fixed anchor $x_0$. The resulting MMD between distributions does not depend on the choice of $x_0$, and $\mathcal{E}(F,G)=2\,\mathrm{MMD}^2_k(F,G)$ \cite[Thm.~22]{Sejdinovic2013}. Consequently, EB is an instance of KOM with kernel $k$ and a variance-regularization parameter set to $\lambda=0$.

\paragraph{Code accessibility.}
Accessible, author-supported implementations are crucial. At the time of our study, EB and KOM lacked clearly linked, official code; \texttt{WeightIt} later provided an EB implementation \cite{WeightItR}, and author reference code facilitated KOM with some reconciliation between code and paper. For TLF, an R source package (\texttt{covalign}, \cite{covalignR}) existed but was not referenced in the article and the recommended routine was not exported, forcing users to inspect source internals. Such barriers raise the risk of user-side re-implementation, bugs, or omission of impactful details (e.g., Gram-matrix diagonalization in TLF).

\paragraph{Weight post-processing.}
We chose not to apply weight post-processing (e.g., truncation \cite{Xiao2013}, trimming \cite{Sturmer2021}, stabilization \cite{Xu2010}) to keep IPTW as a reference and avoid enshrining any single choice across $180{,}000$ datasets. IPTW’s performance has been extensively studied, and careful modeling plus post-processing can work well in practice; our goal was to test newer methods against a simple, transparent baseline. For applied work, we refer readers to practical guidance on diagnostics and post-processing \cite{Austin2015}.

\paragraph{Possible explanations for TLF performance.}
In our study, adopting the author's $\lambda$-selection strategy together with the recommended kernel family, but pairing it with a heuristic bandwidth, did not yield strong results, suggesting that this particular recipe may be mismatched to our design. Other tuning choices discussed by the author could plausibly perform better; however, to our knowledge there is not yet a broadly reliable, data-driven strategy to jointly select the regularization parameter and kernel bandwidth for TLF in finite samples. We highlight the need for future work to develop and validate such joint selection procedures before recommending routine use in settings like ours.

\paragraph{Scope and exclusions.}
We focused on methods capable of estimating ATE and ATT with sufficient detail for reproducible implementation. Other approaches (e.g., entropy balancing, overlap weighting) target different estimands or rely on different design choices and were outside scope. Our goal was not to rank all weighting strategies but to document what contemporary methods deliver under transparent, good-faith use.

\paragraph{Limitations.}
First, we could not fully follow all author-recommended procedures for confidence intervals. For EB, bootstrap CIs were impractical at our simulation scale. For TLF, the honest, design-based CI requires specifying an RKHS and an upper bound on the outcome’s norm, quantities that are not data-verifiable, with limited implementation guidance in low/moderate dimensions.\\
Second, the data-generating mechanism is clinically inspired yet stylized; different outcome structures, higher dimensions, or more severe non-overlap could change relative behavior.

\paragraph{Practical recommendation.}
When empirical overlap is strong and a reasonably specified propensity score model is available, IPTW is a natural and effective choice and is well covered by existing guidance \citep{IPW,IPTW_norm_weight,Austin2015}. By contrast, when overlap is limited or misspecification is plausible, IPTW can become fragile (e.g., extreme weights), often requiring expert interventions such as trimming or alternative targets \citep{IPTW_trim_weight1,Sturmer2021,Lee2011,Li2018}. In such settings, it is useful to triangulate with modern balancing approaches and assess concordance of conclusions. Two practical candidates are EB and KOM: EB can be deployed with few tuning choices and is implemented in \texttt{WeightIt} \citep{Jared2020,WeightItR}, whereas KOM offers an explicit bias-variance trade-off in an RKHS but requires selecting a kernel and a regularization parameter and is typically used via authors' research code \citep{kallusATT,kallus2019optimalATE,kalluscodearticle}. We note that EB relies on Euclidean distances and is therefore scale-dependent; standardization of covariates is advisable \citep{energyDistance}. Although TLF is a promising approach \citep{TLF}, our simulations indicate that adopting the author's $\lambda$-selection strategy and recommended kernel family, with a heuristically chosen bandwidth, did not yield strong finite-sample performance for the treatment-effect estimates. Teams without specialized expertise may wish to prioritize EB (and, when feasible, KOM) for sensitivity analyses or when nonparametric methods are required.

\section{Declaration}

\paragraph{Availability of data and materials.}
R code for data generation and analysis for the current study are available in this article GitHub repository \citep{benchmark_repo}. Docker used to run the simulation available for download at \hyperlink{https://cloud.sylabs.io/library/ep123456/collection/v6}{https://cloud.sylabs.io/library/ep123456/collection/v6}.\\
The \textsc{PROBITsim} dataset \cite{Goetghebeur2020} analysed during the current study are available in the GitHub repository \hyperlink{https://github.com/IngWae/Formulating-causal-questions}{https://github.com/IngWae/Formulating-causal-questions}.
\paragraph{Competing interests.}
The authors declare that they have no competing interests.
\paragraph{Funding.}
Etienne Peyrot acknowledges support from the Université Paris Cité. Francois Petit acknowledges support from the French Agence Nationale de la Recherche through the project reference ANR-22-CPJ1-0047-01. Rapha\"el Porcher acknowledges support from the French Agence Nationale de la Recherche as part of the “Investissements d’avenir” program, reference ANR-23-IACL-0008  (PR[AI]RIE-PSAI IA cluster). This work was partially funded by the Agence Nationale de la Recherche, under grant agreement no. ANR-18-CE36-0010-01.
\paragraph{Authors' contributions.}
All authors were involved in the study concept and design, the analysis and interpretation of the data and, the drafting of the manuscript. EP did the code implementation, the figures and the tables.
\paragraph{Acknowledgements.}
Numerical computations were partly performed on the S-CAPAD/DANTE platform, IPGP, France.


\begin{thebibliography}{10}

\bibitem{IPW}
Rosenbaum PR, Rubin DB.
\newblock The central role of the propensity score in observational studies for
  causal effects.
\newblock Biometrika. 1983 04;70(1):41-55.

\bibitem{IPTW_norm_weight}
Lunceford JK, Davidian M.
\newblock Stratification and weighting via the propensity score in estimation
  of causal treatment effects: a comparative study.
\newblock Stat Med. 2004 Oct;23(19):2937-60.

\bibitem{Hahn1998}
Hahn J.
\newblock On the Role of the Propensity Score in Efficient Semiparametric
  Estimation of Average Treatment Effects.
\newblock Econometrica. 1998;66(2):315-31.

\bibitem{Smit2023}
Smit JM, Krijthe JH, Kant WMR, Labrecque JA, Komorowski M, Gommers DAMPJ,
  et~al.
\newblock Causal inference using observational intensive care unit data: a
  scoping review and recommendations for future practice.
\newblock npj Digital Medicine. 2023;6(1):221.

\bibitem{Zuo2023}
Zuo H, Yu L, Campbell SM, Yamamoto SS, Yuan Y.
\newblock The implementation of target trial emulation for causal inference: a
  scoping review.
\newblock Journal of Clinical Epidemiology. 2023;162:29-37.

\bibitem{AIPW1}
Robins JM, Rotnitzky A, Zhao LP.
\newblock Estimation of Regression Coefficients When Some Regressors are not
  Always Observed.
\newblock Journal of the American Statistical Association. 1994;89(427):846-66.

\bibitem{IPW1952}
Horvitz DG, Thompson DJ.
\newblock A Generalization of Sampling Without Replacement From a Finite
  Universe.
\newblock Journal of the American Statistical Association. 1952;47(260):663-85.

\bibitem{Jared2020}
Huling JD, Mak S.
\newblock Energy balancing of covariate distributions.
\newblock Journal of Causal Inference. 2024;12(1):20220029.

\bibitem{kallusATT}
Kallus N.
\newblock Generalized Optimal Matching Methods for Causal Inference.
\newblock Journal of Machine Learning Research. 2020;21(62):1-54.

\bibitem{kallus2019optimalATE}
Kallus N, Santacatterina M.
\newblock Optimal Estimation of Generalized Average Treatment Effects using
  Kernel Optimal Matching.
\newblock arXiv preprint arXiv:190804748. 2019.

\bibitem{TLF}
Zhao Q.
\newblock Covariate balancing propensity score by tailored loss functions.
\newblock The Annals of Statistics. 2019;47(2):965-93.

\bibitem{Hainmueller2012}
Hainmueller J.
\newblock Entropy Balancing for Causal Effects: A Multivariate Reweighting
  Method to Produce Balanced Samples in Observational Studies.
\newblock Political Analysis. 2012;20(1):25-46.

\bibitem{Franklin2013}
Franklin JM, Rassen JA, Ackermann D, Bartels DB, Schneeweiss S.
\newblock Metrics for covariate balance in cohort studies of causal effects.
\newblock Statistics in Medicine. 2013;33(10):1685-99.

\bibitem{Rubin1974}
Rubin DB.
\newblock Estimating causal effects of treatments in randomized and
  nonrandomized studies.
\newblock Journal of Educational Psychology. 1974;66(5):688-701.

\bibitem{Neyman1923}
Neyman J.
\newblock On the application of probability theory to agricultural experiments.
  Essay on Principles. Section 9 (translation published in 1990).
\newblock Statistical Science. 1923;5:472-80.

\bibitem{hajek1971}
H{\'a}jek J.
\newblock Comment on "An Essay on the Logical Foundations of Survey Sampling,
  Part One" by D. Basu.
\newblock In: Godambe VP, Sprott DA, editors. Foundations of Statistical
  Inference. Toronto: Holt, Rinehart and Winston; 1971. p. 236-48.

\bibitem{AIPW2}
Scharfstein DO, Rotnitzky A, Robins JM.
\newblock Adjusting for Nonignorable Drop-Out Using Semiparametric Nonresponse
  Models.
\newblock Journal of the American Statistical Association.
  1999;94(448):1096-120.

\bibitem{Tsiatis2006}
Tsiatis A.
\newblock Semiparametric Theory and Missing Data. vol.~73.
\newblock Springer New York; 2006.

\bibitem{newey1994}
Newey WK, McFadden DL.
\newblock Chapter 36 Large sample estimation and hypothesis testing.
\newblock In: Handbook of Econometrics. vol.~4 of Handbook of Econometrics.
  Elsevier; 1994. p. 2111-245.

\bibitem{Vaart1998}
van~der Vaart AW.
\newblock Asymptotic Statistics.
\newblock Cambridge Series in Statistical and Probabilistic Mathematics.
  Cambridge University Press; 1998.

\bibitem{StefanskiBoos2002}
Stefanski LA, Boos DD.
\newblock The {C}alculus of {M}-{E}stimation.
\newblock The American Statistician. 2002;56(1):29-38.

\bibitem{energyDistance}
Sz{\'e}kely GJ, Rizzo ML.
\newblock Energy statistics: A class of statistics based on distances.
\newblock Journal of Statistical Planning and Inference. 2013;143(8):1249-72.

\bibitem{Freedman2006}
Freedman DA.
\newblock On The So-Called “Huber Sandwich Estimator” and “Robust
  Standard Errors”.
\newblock The American Statistician. 2006;60(4):299-302.

\bibitem{Kang2007}
Kang JDY, Schafer JL.
\newblock Demystifying Double Robustness: A Comparison of Alternative
  Strategies for Estimating a Population Mean from Incomplete Data.
\newblock Statistical Science. 2007;22(4).

\bibitem{benchmark_repo}
Peyrot E.
\newblock Weighting methods pragmatical review: Analysis code and materials.
\newblock GitHub; 2025.
\newblock Available from:
  \url{https://github.com/EtiennePeyrot/benchmark_balancing_methods}.

\bibitem{kalluscodearticle}
Kallus N, Pennicooke B, Santacatterina M.
\newblock More robust estimation of sample average treatment effects using
  Kernel Optimal Matching in an observational study of spine surgical
  interventions.
\newblock arXiv preprint arXiv:181104274. 2018.
\newblock Code: \url{https://github.com/CausalML/KOM-SATE}.

\bibitem{gurobi}
{Gurobi Optimization, LLC}.
\newblock Gurobi Optimizer Reference Manual.
\newblock Gurobi Optimization, LLC; 2024.
\newblock Available from: \url{https://www.gurobi.com}.

\bibitem{osqp}
Stellato B, Banjac G, Goulart P, Boyd S.
\newblock OSQP: Quadratic Programming Solver using the OSQP Library. R package
  version 0.6.0.5.
\newblock CRAN; 2021.
\newblock Available from: \url{https://CRAN.R-project.org/package=osqp}.

\bibitem{covalignR}
Zhao Q.
\newblock covalign: R source code for covariate balancing by tailored loss
  functions (CBSR/TLF).
\newblock University of Cambridge; 2019.
\newblock Available from:
  \url{https://www.statslab.cam.ac.uk/~qz280/publication/balancing-loss/}.

\bibitem{gretton12}
Gretton A, Borgwardt KM, Rasch MJ, Sch{\"o}lkopf B, Smola A.
\newblock A Kernel Two-Sample Test.
\newblock Journal of Machine Learning Research. 2012;13(25):723-73.

\bibitem{garreau2018}
Garreau D, Jitkrittum W, Kanagawa M.
\newblock Large Sample Analysis of the Median Heuristic.
\newblock arXiv preprint arXiv:170707269. 2017.

\bibitem{Ramdas2015}
Ramdas A, Reddi SJ, P{\'o}czos B, Singh A, Wasserman L.
\newblock On the Decreasing Power of Kernel and Distance Based Nonparametric
  Hypothesis Tests in High Dimensions.
\newblock Proceedings of the AAAI Conference on Artificial Intelligence. 2015
  Mar;29(1).

\bibitem{Schrab2023}
Schrab A, Kim I, Albert M, Laurent B, Guedj B, Gretton A.
\newblock {MMD} Aggregated Two-Sample Test.
\newblock Journal of Machine Learning Research. 2023;24(194):1-81.

\bibitem{WeightItR}
Greifer N.
\newblock WeightIt: Weighting for Covariate Balance in Observational Studies. R
  package version 0.13.1.
\newblock CRAN; 2022.
\newblock Available from: \url{https://CRAN.R-project.org/package=WeightIt}.

\bibitem{R}
{R Core Team}.
\newblock R: A Language and Environment for Statistical Computing.
\newblock Vienna, Austria: R Foundation for Statistical Computing; 2022.
\newblock Available from: \url{https://www.R-project.org/}.

\bibitem{Goetghebeur2020}
Goetghebeur E, {le Cessie} S, {De Stavola} B, Moodie EEM, Waernbaum I.
\newblock Formulating causal questions and principled statistical answers.
\newblock Statistics in Medicine. 2020;39(30):4922-48.

\bibitem{Sejdinovic2013}
Sejdinovic D, Sriperumbudur B, Gretton A, Fukumizu K.
\newblock Equivalence of distance-based and {RKHS}-based statistics in
  hypothesis testing.
\newblock The Annals of Statistics. 2013;41(5).

\bibitem{Xiao2013}
Xiao Y, Moodie EEM, Abrahamowicz M.
\newblock Comparison of Approaches to Weight Truncation for Marginal Structural
  Cox Models.
\newblock Epidemiologic Methods. 2013;2(1):1-20.

\bibitem{Sturmer2021}
St{\"u}rmer T, Webster-Clark M, Lund JL, Wyss R, Ellis AR, Lunt M, et~al.
\newblock Propensity Score Weighting and Trimming Strategies for Reducing
  Variance and Bias of Treatment Effect Estimates: A Simulation Study.
\newblock American Journal of Epidemiology. 2021;190(8):1659-70.

\bibitem{Xu2010}
Xu S, Ross C, Raebel MA, Shetterly S, Blanchette C, Smith D.
\newblock Use of Stabilized Inverse Propensity Scores as Weights to Directly
  Estimate Relative Risk and Its Confidence Intervals.
\newblock Value in Health. 2010;13(2):273-7.

\bibitem{Austin2015}
Austin PC, Stuart EA.
\newblock Moving towards best practice when using inverse probability of
  treatment weighting ({IPTW}) using the propensity score to estimate causal
  treatment effects in observational studies.
\newblock Statistics in Medicine. 2015;34(28):3661-79.

\bibitem{IPTW_trim_weight1}
Crump RK, Hotz VJ, Imbens GW, Mitnik OA.
\newblock {Dealing with limited overlap in estimation of average treatment
  effects}.
\newblock Biometrika. 2009;96(1):187-99.

\bibitem{Lee2011}
Lee BK, Lessler J, Stuart EA.
\newblock Weight Trimming and Propensity Score Weighting.
\newblock PLoS ONE. 2011;6(3):e18174.

\bibitem{Li2018}
Li F, Thomas LE.
\newblock Addressing Extreme Propensity Scores via the Overlap Weights.
\newblock American Journal of Epidemiology. 2018.

\end{thebibliography}
\end{document}